\documentclass[12pt]{amsart}

\usepackage{amssymb}

\usepackage{euscript}
\usepackage{multicol}
\usepackage{amsmath}
\usepackage{times}

\date{}

\begin{document}

\title[Monotonicity of  the set of zeros ]{{\vspace*{-1.5cm}\sc Monotonicity of  the set of zeros of the Lyapunov exponent with respect  to  shift  embeddings}}

\author{Oleg Safronov}
\email{osafrono@uncc.edu}
\address{Department of Mathematics and Statistics,  UNCC,   Charlotte,  NC}

\maketitle

\begin{abstract}We consider  the discrete Schr\"odinger operators
with  potentials whose    values are   read along the orbits of a shift  of  finite  type. 
We   study  a  certain subset  of the  collection of energies  at  which  the Lyapunov exponent is  zero and prove 
monotonicity   of this set with respect to  the   shift embeddings.  Then we  introduce a  certain function ${\mathcal J}(A,\mu)$  determined  by the position of these  zeros
and 
prove  monotonicity  of ${\mathcal J}(A,\mu)$ with respect  to  embeddings.
\end{abstract}






\newtheorem{theorem}{Theorem}
\newtheorem{lemma}[theorem]{Lemma}
\newtheorem{proposition}[theorem]{Proposition}
\newtheorem{example}[theorem]{Example}
\newtheorem{definition}[theorem]{Definition}
\newtheorem{corollary}[theorem]{Corollary}
\newtheorem{remark}[theorem]{Remark}
\newtheorem{assumption}[theorem]{Assumption}
\newtheorem{conjecture}[theorem]{Conjecture}
\newtheorem{claim}[theorem]{Claim}

\renewcommand{\thefootnote}{\ensuremath{\fnsymbol{footnote}}}

\def\theequation{\arabic{section}.\arabic{equation}}

\def\define{\stackrel{\Delta}{=}}
\def\real{\mathbb{R}}
\def\Real{\mathbb{R}}
\def\R{\mathbb{R}}
\def\endproof{\hfill\diamond}
\def\sB{{\cal B}}
\def\sF{{\cal F}}
\def\sG{{\cal G}}
\def\sf{{\cal F}}
\def\sg{{\cal G}}
\def\sH{{\cal H}}
\def\sL{{\cal L}}
\def\sh{{\cal H}}
\def\sK{{\cal K}}
\def\sC{{\cal C}}
\def\sQ{{\cal Q}}
\def\Prob{\mathbb{P}}
\def\Prob{\mathbb{R}}
\def\Expn{\mathbb{E}}
\def\P{\mathbb{P}}
\def\E{\mathbb{E}}
\def\Ind{I}
\def\ind{I}
\def\Wtilde{\widetilde{W}}
\def\Btilde{\widetilde{B}}
\def\Ptilde{\widetilde{\Prob}}
\def\Etilde{\widetilde{\Expn}}
\def\Ltilde{\widetilde{L}}
\def\ptilde{\tilde{p}}
\def\qtilde{\tilde{q}}
\def\Ntilde{\widetilde{N}}
\def\ntilde{\tilde{n}}
\def\btilde{\tilde{b}}
\def\sgn{\mbox{\rm sgn}}

\bigskip

In this short paper,  we study  the discrete Schr\"odinger
operators $H_\omega$ defined   on $\ell^2({\Bbb Z})$  by
$$
\bigl[H_\omega u \bigr](n)=u(n+1)+u(n-1) +V(T^n \omega) u(n),\qquad \omega\in \Omega.
$$
Here $\Omega$  is  a compact metric  space whose elements are  infinite sequences $\{\omega_n\}_{n\in {\Bbb Z}}$  such that  $\omega_n\in \{1,\dots,\ell\}={\mathcal A}$ for each $n$.
There  are sequences in ${\mathcal A}^{\Bbb Z}$  that are not allowed to be in $\Omega$ and we assume that forbidden words are of length $2$.
The   metric $d(\cdot,\cdot)$  on $\Omega$  is  defined by  $$d(\omega,\omega')=e^{-N(\omega,\omega')},$$
where $N(\omega,\omega')$ is  the  largest  nonnegative integer   such that $\omega_n=\omega_n'$   for all $|n|< N(\omega,\omega')$.
The mapping      $T:\,\Omega\to\Omega$  is  assumed  to be  a subshift of  finite type defined by
$$
\bigl(T\omega\bigr)_n=\omega_{n+1},\qquad  \forall n\in {\Bbb Z}.
$$
Finally,  the function $V$  is  assumed to be  locally  constant on $\Omega$ in the sense of the following definition.

{\it Definition}.  A  function $V:\, \Omega\to {\Bbb R} $  is said to be locally constant,   if there is an $\epsilon>0$  such that
$$
V(\omega')=V(\omega)\qquad  \text{whenever}\qquad d(\omega',\omega)<\epsilon.
$$

\bigskip

Spectral properties of $H_\omega$ are related to the 
behavior of  solutions to the equation
\begin{equation}\label{SEquation}
u(n+1)+u(n-1) +V(T^n \omega) u(n)=Eu(n),\qquad n\in {\Bbb Z},
\end{equation}  for $E\in {\Bbb R}$.

On the other hand,   all  solutions
to \eqref{SEquation} can be described in terms of the Schr\"odinger cocycles $(T, A^E)$ with  $A=A^E:\, \Omega\to{\rm SL}(2, {\Bbb R})$ defined by
$$
A^E(\omega)=\begin{pmatrix}
E-V(\omega)& -1\\
1& 0
\end{pmatrix}
$$
Namely,  $u$ is a solution of \eqref{SEquation} if  and only if
$$
\begin{pmatrix} u(n)\\ u(n-1) \end{pmatrix}= A_n(\omega)\cdot  \begin{pmatrix} u(0)\\ u(-1) \end{pmatrix},\qquad \forall n\in {\Bbb Z},
$$
where
$$
A_n(\omega)=\begin{cases}A(T^{n-1}\omega)\cdots A(\omega)\quad \text{if}\quad n\geq 1;\\
[A_{-n}(T^n\omega)]^{-1}
\quad \text{if}\quad n\leq -1;\\
{\rm Id}\quad \text{if}\quad n=0.
\end{cases}
$$

Since  $\Omega$  is a metric space,   we  can talk about 
 the Borel $\sigma$-algebra of  subsets of $\Omega$  and consider probability measures on $\Omega$.
Let $\mu$  be a $T$-ergodic  probability   measure on $\Omega$. The Lyapunov exponent  for $A$ and $\mu$
is defined by
$$
L(A,\mu)=\lim_{n\to\infty}\frac1n \int \ln(\|A_n(\omega)\|)d\mu(\omega).
$$
By Kingman’s subaddive ergodic theorem, 
$$
\frac1n \ln(\|A_n(\omega)\|)\qquad \text{converges to}\quad L(A,\mu)\qquad \text{as}\quad n\to \infty,
$$
for $\mu$-almost every $\omega\in \Omega$.  For  simplisity,  we  write $L(E)=L(A,\mu).$

One of the  main theorems  of  the paper \cite{ADZ}  gives  sufficient   conditions guaranteeing   that  the set
\begin{equation} \label{DefL} {\frak L}(A,\mu)=\bigl\{E\in {\Bbb R}:\,\, L(A,\mu)=0\bigr\}
\end{equation}
is  finite.  One of these conditions is   that $\mu$ has  a local product structure.

Let us now give a  formal  definition of  a measure  having  this  property.
We first define  the   spaces  of semi-infinite  sequences
$$
\Omega_+=\{\{\omega_n\}_{n\geq0}:\,\, \omega\in \Omega\}
\quad\text{
and 
}\quad
\Omega_-=\{\{\omega_n\}_{n\leq0}:\,\, \omega\in \Omega\}.
$$
Then using  the natural projection  $\pi_\pm$
 from $\Omega$ onto $\Omega_\pm$,  we  define  $\mu_\pm=(\pi_\pm)_*\mu$  on $\Omega_\pm$  to
 be
the pushforward measures of $\mu$.  After  that,  for each $1\leq j\leq \ell$, we introduce
 the
cylinder sets
$$
[0;j]=\{\omega\in\Omega:\,\, \omega_0=j\}\quad
\text{and}\quad
[0;j]_\pm=\{\omega\in\Omega_\pm:\,\, \omega_0=j\}.
$$ 
A local product structure  is  a relation between   the measures $\mu_j=\mu\bigl|_{[0;j]}$  and   the measures $\mu_j^\pm=\mu_\pm\bigl|_{[0;j]}$.
To describe this relation, we need  to consider
the 
natural homeomorphisms
$$P_j: [0;j]\to [0;j]_-\times [0;j]_+   $$ 
defined  by  $$P_j(\omega)=\bigl(\pi_-\omega,\pi_+\omega\bigr),\qquad \forall \omega\in \Omega.$$

\bigskip

{\it Definition}. We say that $\mu$  has a local product structure if there is a positive   $\psi:\Omega\to (0,\infty)$
such that for each $1\leq j\leq \ell$, the function  $\psi\circ P_j^{-1}$  belongs to $ L^1\bigl( [0;j]_-\times [0;j]_+, \mu^-_j\times\mu^+_j\bigr)$
and
$$
\bigl(P_j\bigr)_*d\mu_j=\psi\circ P_j^{-1}\, d( \mu^-_j\times\mu^+_j).
$$

\bigskip

We will shortly divide 
points  of the set \eqref{DefL} into two  groups: removable and  unremovable  points.  We  will show  that unremovable  points in ${\frak L}(A,\mu)$
do not disappear  in the process  of  passing    from  $\Omega$  to a  subshift  $\tilde \Omega\subset \Omega$  with  an ergodic measure $\tilde \mu$ on it.

A point $p\in \Omega$  is said to be  periodic for $T$ provided  there is  a positive integer $n_p$
for which $T^{n_p}\,p=p$.
If  $p\in \Omega $   is  periodic,  then  $V(T^np)$  is a periodic function of $n$,  because 
$V(T^{n_p + n}p) = V(T^np)$ for every $n\in {\Bbb Z}$.  
For a periodic   point  $p$ of period $n_p$, define  $\Delta_p(E)$ to be the trace of  the monodromy matrix $A_{n_p}(p)$ 
$$ \Delta_p(E) ={ \rm Tr}(A_{n_p}(p)).$$
By   ${\rm Per}(T)$,  we denoted  the collection of all periodic points of $T$.

{\it Definition}.  A point $E\in {\frak L}(A,\mu)$  is said  to be unremovable from ${\frak L}(A,\mu)$  provided  

either $\,\,\,$  1)   there  exists a $T$- periodic point  $p\in \Omega$  for which $0<|\Delta_p(E)|<2$,

  or  $\qquad $  2)  
$|\Delta_p(E)|\in\{0,2\}$  for all  $p\in {\rm Per}(T)$.

The collection of  unremovable  from ${\frak L}(A,\mu)$  points   will be  denoted by ${\frak U}(A,\mu)$.

\bigskip

\begin{theorem}\label{thm2}  Let $T:\, \Omega\to\Omega$ be  a  subshift of finite type.
Assume that $\mu$ is a $T$-ergodic  measure on $\Omega$ that  has a local product  structure and   the property ${\rm supp}(\mu)=\Omega$.
Let $V$ be a  real-valued   locally  constant function on $\Omega$.  
Then   for any   subshift $\tilde T:\, \tilde\Omega\to\tilde \Omega$   of $T$
  and  any 
$\tilde T$-ergodic measure $\tilde \mu$ on  $\tilde \Omega\subset \Omega$,
\begin{equation}\label{U}
  {\frak U}(A,\mu)\,\subseteq  {\frak U}(\tilde A,\tilde \mu),
\end{equation}
where $\tilde A$ is the restriction of $A$  to $\tilde \Omega$.
\end{theorem}

\bigskip

\begin{corollary}\label{c2}  Let $T:\, \Omega\to\Omega$ be  a subshift of finite type.
Assume that $\mu$ and $\tilde \mu$  are $T$-ergodic  measures on $\Omega$ that  have a local product  structure  and  have the property ${\rm supp}(\mu)={\rm supp}(\tilde \mu)=\Omega$.
Let $V$ be a  real-valued   locally  constant function on $\Omega$.  
Then  
\begin{equation}\label{U2}
  {\frak U}(A,\mu)\,= {\frak U}( A,\tilde \mu).
\end{equation}
\end{corollary}

\bigskip

\noindent
{\bf Remark}.  For   any removable  point $E$  that belongs to  the set ${\frak L}(A,\mu)\setminus {\frak U}(A,\mu)$, there   is a    subshift  $\tilde T:\, \tilde\Omega\to\tilde \Omega$   of $T$
 and  a 
$\tilde T$-ergodic measure $\tilde \mu$ on  $\tilde \Omega\subset \Omega$   for which $E\notin {\frak L}(\tilde A,\tilde \mu)$.  To see that, we  find a periodic point $p\in \Omega$  for which $|\Delta_p(E)|>2$
and then we  define $\tilde \mu$  to be   the  ergodic probability  measure  supported  on the    union of the shifts $T^np$  of the  point $p$.

\bigskip

The following  result is  a  consequence of our methods:

\begin{theorem} Let $T:\, \Omega\to\Omega$ be  a  subshift of finite type.
Assume that $\mu$ is a $T$-ergodic  measure on $\Omega$ that  has a local product  structure and   the property ${\rm supp}(\mu)=\Omega$.
Let $V$ be a  real-valued   locally  constant function on $\Omega$.  
Then   
\begin{equation}\notag
  {\frak U}(A,\mu)\,=\bigcap_{p\in {\rm Per}(T)} \sigma(p),
\end{equation}
where $\sigma(p)$ denotes   the spectrum of the  Schr\"odinger operator $H_p$  with  the potential  $V(T^n p)$.
\end{theorem}

Several  definitions  below   ivolve the  set
\[
{\frak S}(T,\mu)= \bigcup_{p\in {\rm Per}(T)} \bigl\{E\in {\Bbb R}:\,\,\Delta_p(E)\in (-2,0)\cup(0,2)\bigr\}.\]
This set  may only become  smaller  when one  passes    from $T$  to   $\tilde T$,
$$
{\frak S}(\tilde T,\tilde \mu)\subseteq {\frak S}(T,\mu),
$$
while  $ {\frak U}(A,\mu)$  may only increase   due to  the   property \eqref{U}.  This  observation allows  one   to  construct a  real-valued  function  ${\mathcal  J}(A,\mu)$
that decreases  when  either   $ {\frak S}(T,\mu)$  becomes smaller,   or  $ {\frak U}(A,\mu)$  becomes  larger.
For this  purpose,  we  recall  that if $T$  has a fixed point,   then  there  are  at most  finitely many points in the set ${\frak U}(A, \mu)$ (see Theorem 1.2  in \cite{ADZ}).   Thus ${\frak U}(A, \mu)=\{E_1,E_2,\dots,E_l\}$   where    $E_1<E_2<\dots<E_{l}$.
We  first enlarge the  collection ${\frak U}(A, \mu)$  by adding the two points $E_0=-5/2-\|V\|_\infty$  and $E_{l+1}=5/2+\|V\|_\infty$. Then,
for each interval $(E_j,E_{j+1})$ whose intersection with ${\frak S}(T,\mu)$ is not empty,  we define $N_j$  by   
$$
N_j=\text{ the integer part of   }\Bigl[\frac{2|E_{l+1}-E_0|}{|(E_j,E_{j+1})\cap {\frak S}(T,\mu)\bigr|}\Bigr].
$$
Here, $|X|$  in the denominator denotes the Lebesgue measure of a  Borel set $X\subset {\Bbb R}$.   If $(E_j,E_{j+1})\cap {\frak S}(T,\mu)=\emptyset$, then  we    define $N_j$  to be equal to $2$. 
Finally,  after setting   
$$
N(A, \mu)=\max_{0\leq j \leq l}N_j,
$$
we define  the function 
$$
{\mathcal  J}(A,\mu)=\sum_{j=0}^{l} {\mathcal   E}_{j}(A,\mu),
$$
where   
\[\begin{split}
 {\mathcal   E}_{j}(A,\mu)=\frac{\bigl|(E_j,E_{j+1})\cap {\frak S}(T,\mu)\bigr|}{\lambda}\,\, \ln\Bigl(\frac{\bigl|(E_j,E_{j+1})\cap {\frak S}(T,\mu)\bigr|}{\lambda}\Bigr)+\\+
\frac{\bigl|(E_0,E_{l+1})\setminus {\frak S}(T,\mu)\bigr|}{\lambda\cdot (l+1)}\,\, \ln\Bigl(\frac{\bigl|(E_0,E_{l+1})\setminus {\frak S}(T,\mu)\bigr|}{N(A,\mu)\cdot \lambda}\Bigr)
\end{split}
\]   and $\lambda=|E_{l+1}-E_0|$.
The next  result  establishes  monotonicity  of   the   function ${\mathcal  J}(A,\mu)$  with respect to  embeddings of the subshifts.

\begin{theorem}\label{thm3} Let $T:\, \Omega\to\Omega$   be  a subshift of finite type.
Assume that $\mu$ is a $T$-ergodic  measure on $\Omega$ that  has a local product  structure  and the property ${\rm supp}(\mu)=\Omega$.
 Let $V$ be a  real-valued   locally  constant function on $\Omega$.  
Suppose $\tilde T:\, \tilde\Omega\to\tilde \Omega$  is a  further subshift   of $T$
  and $\tilde \mu$  is a
$\tilde T$-ergodic measure on  $\tilde \Omega\subset \Omega$  for which    the set ${\frak U}(\tilde A,\tilde \mu)$ is finite  (by $\tilde A$,  we denote  the restriction of $A$  to $\tilde \Omega$).  Then
$$
 {\mathcal  J}(A,\mu)\,\geq  {\mathcal  J}(\tilde A,\tilde \mu).
$$

\end{theorem}

\bigskip

\section{Begining of the proof of Theorem~\ref{thm2}.  Main ingredients}

Note  that
 a Schr\"odinger cocycle $A=A^E$ with a locally constant potential $V:\Omega\to {\Bbb R}$ is  also locally constant.
Put differently,  there is an $\epsilon>0$  such that
$$
A(\omega')=A(\omega)\qquad  \text{whenever}\qquad d(\omega',\omega)<\epsilon.
$$

{\it Definition }.   Let $T:\Omega\to \Omega$ be a subshift of finite type.
The local stable set of a point $\omega\in \Omega$ is defined by
$$
W^s(\omega)=\{\omega'\in\Omega:\,\, \omega'_n=\omega_n\quad \text{for}\quad n\geq 0\}
$$
and the local unstable set of $\omega$  is defined by
$$
W^u(\omega)=\{\omega'\in\Omega:\,\, \omega'_n=\omega_n\quad \text{for}\quad n\leq 0\}.
$$

\bigskip

For $\omega'\in W^s(\omega)$,  define $H_{\omega',\omega}^{s,n}$  to be
$$
H_{\omega,\omega'}^{s,n}=\bigl[A_n(\omega')\bigr]^{-1}A_n(\omega).
$$
Since $d(T^j\omega',T^j\omega)\leq e^{-j}$ tends to $0$ as $j\to \infty$,  there is an index $n_0$  for which
$$
H_{\omega,\omega'}^{s,n}=H_{\omega,\omega'}^{s,n_0}\qquad \text{for }\quad  n\geq n_0.
$$
In this case, we define  the stable  holonomy  $H_{\omega,\omega'}^s$ by
$$
H_{\omega,\omega'}^s=H_{\omega,\omega'}^{s,n_0}.
$$
The unstable holonomy $H_{\omega,\omega'}^u$  for $\omega'\in W^u(\omega)$   is  defined similarly by
$$
H_{\omega,\omega'}^u=
\bigl[A_n(\omega')\bigr]^{-1}A_n(\omega)   \qquad \text{for all}\quad n\leq -n_0.
$$

The general theory of dynamical systems  tells us that
the  cocycle
$$
(T,A): \Omega\times {\Bbb R \Bbb P}^1\to {\Bbb R \Bbb P}^1
$$
defined  by 
$$
(T, A)(\omega,\xi)= (T\omega, A(\omega) \xi)
$$
has an invariant probability measure $m$  on $\Omega\times {\Bbb R \Bbb P}^1.$ We say that such a measure $m$ projects to $\mu$
if $m(\Delta \times {\Bbb R \Bbb P}^1)=\mu(\Delta)$  for all Borel subsets $\Delta$ of $\Omega$.  Given any $T$-invariant  measure $\mu$  on $\Omega$,
   one can  find a $(T,A)$-ivariant measure  $m$  that projects to $\mu$ by applying  the Krylov-Bogolyubov trick.

\smallskip

{\it Definition}.  Suppose $m$ is  a $(T, A)$-invariant probability measure 
on $\Omega\times {\Bbb R \Bbb P}^1$
that projects to $\mu$.   A disintegration of $m$ 
 is a measurable family $\{m_\omega:\quad \omega\in \Omega\}$ of probability measures on ${\Bbb R \Bbb P}^1$
having the property
$$
m(D)=\int_\Omega m_\omega(\{\xi\in {\Bbb R \Bbb P}^1:\,\, (\omega,\xi)\in D\})d\mu(\omega)
$$
for each measurable set  $D\subset \Omega\times{\Bbb R \Bbb P}^1.$

\smallskip
  Existence of  such a disintegration is guaranteed 
by Rokhlin’s theorem. 
Moreover,
$\{\tilde m_\omega:\quad \omega\in \Omega\}$  is another disintegration of $m$ then  $m_\omega=\tilde m_\omega$  for $\mu$-almost
every $\omega\in \Omega$. 
It is  easy to see  that $m$  is
$(T, A)$-invariant if and only if $A(\omega)_*m_\omega=m_{T\omega}$
for $\mu$-almost every  $\omega\in \Omega$.

\smallskip

{\it Definition}.  
A $(T,A)$-invariant measure $m$  on $\Omega\times {\Bbb R \Bbb P}^1$ that projects to $\mu$ is said to be an su-state for $A$ provided it has 
a disintegration $\{m_\omega:\quad \omega\in \Omega\}$  such that   for  $\mu$-almost every $\omega\in \Omega $, 
$$\,$$
1)$$\quad A(\omega)_* m_\omega=m_{T \omega},$$
2) $$\bigl( H^s_{\omega,\omega'}\bigr)_*m_\omega=m_{\omega'}\qquad
\text{ for every}\quad \omega'\in W^s(\omega).$$
3)$$ \bigl( H^u_{\omega,\omega'}\bigr)_*m_\omega=m_{\omega'}\qquad
\text{ for every}\quad \omega'\in W^u(\omega)$$

\smallskip

The following statement  was  proved in \cite{ADZ}  (Proposition 4.7)  for a   significantly  larger  class of functions $A$.
\begin{proposition} Let $A$ be locally constant.  Suppose $\mu$ has a local product structure   and $L(A,\mu)=0$.   If the support of   the measure  $\mu$  coincides  with all of $\Omega$,  then there exists an su-state for $A$.
\end{proposition}
We apply the  following method to extend $m_\omega$ to a continuous function of $\omega$ on all of $\Omega$.
For each $1\leq j\leq \ell$,  we select     a  point   $\omega^{(j)}$  in $[0;j]\cap \Omega_0$ for which  the measure $m_{\omega^{(j)}}$  is well defined.
Then we set
\begin{equation}\label{definem}
m_{\omega}=\Bigl( H^u_{\omega\wedge \omega^{(\omega_0)}, \omega }   H^s_{\omega^{(\omega_0)}, \omega\wedge \omega^{(\omega_0)}}\Bigr)_*  m_{\omega^{(\omega_0)}}.
\end{equation}
Obviously  $m_\omega$ depends  continuously on $\omega$.

\bigskip

Observe  that ${\Bbb R \Bbb P}^1$ may  be   aslo viewed  as ${\Bbb R}\cup \{ \infty\}$,
because  any vector of the form  $(\xi, 1)\in {\Bbb R \Bbb P}^1$  is uniquily  characterized by $\xi\in {\Bbb R}\cup \{\infty\}$.
Aslo,  ${\Bbb C \Bbb P}^1$ may  be  aslo viewed  as ${\Bbb C}\cup \{ \infty\}$  because  there is a  1:1    mapping of  one set   onto another.
The  part  of  ${\Bbb C \Bbb P}^1$  that is  mapped onto the extended upper half-plane ${\Bbb C}_+\cup \{ \infty\}$   will be denoted  by ${\Bbb C}_+{ \Bbb P}^1$.

Now we  will  state Proposition 4.9  from \cite{ADZ}  in the following more convenient form:
\begin{proposition}\label{barycenter} For each probability measure $\nu$ on ${\Bbb R \Bbb P}^1$
containing
no atom of mass $\geq 1/2$, there is an unique point  $B(\nu)\in {\Bbb C_+ \Bbb  P}$, called the conformal
barycenter of $\nu$, such that 
$$
B(P_* \nu) = P\cdot B(\nu)
$$
 for each $P\in {\rm SL}(2,{\Bbb R})$.
\end{proposition}

Let $m$  be an su-state   with a continuous  disintegration $m_\omega$.
If $m_\omega$   does not have an  atom  of mass $\geq 1/2$, then we set $Z(\omega)\subset {\Bbb C_+ \Bbb  P}$ to   be $\{B(m_\omega)\}$.
Otherwise $Z(\omega)$ is defined to be the   collection of points $\xi $ at  which $m_\omega(\{\xi\})\geq 1/2$.
Since $m_\omega$ is a probability measure,  the set $Z(\omega)$  can  contain at most  two points.
The following  theorem is a consequence of Proposition~\ref{barycenter}.

\begin{theorem} Let $A$ be locally constant.  Suppose $\mu$ has a local product structure   and $L(A,\mu)=0$.   Then
$$
A(\omega) Z(\omega)=Z(T\omega) \qquad  \text{for each}\quad \omega\in \Omega.
$$
 If $\omega',\omega$  are two points in $\Omega$  such that  $\omega_0'=\omega_0$,
then
\begin{equation}\label{Zinvarient}
Z(\omega)=\Bigl( H^u_{\omega\wedge \omega', \omega }   H^s_{\omega', \omega\wedge \omega'}\Bigr) Z(\omega').
\end{equation}
In particular, the number of the points in $Z(\omega)$  does not  depend on $\omega$.   Moreover,  if $Z(\omega)$ is real  for one $\omega$, then it is   real for all $\omega\in \Omega$.
\end{theorem}

The  last two lines of the theorem follow  from the fact  that    for any two points $\omega$  and $\omega'$ in $\Omega$, there is a  real matrix $P\in {\rm SL}(2,{\Bbb R})$
for which  $Z(\omega)=P\cdot  Z(\omega')$.   Indeed, if $\omega'_0=\omega_0$, then this property is  guaranteed  by  \eqref{Zinvarient}.   On the other hand, since $T$ is transitive,
for any two points $\omega'$ and $\omega$,
there is an index $n$  and a point $\tilde \omega$ such that $(T^n\tilde\omega\bigr)_0= \omega'_0$  while  $\tilde \omega_0=\omega_0$.  Therefore
$$
Z(T^n\tilde \omega)=A_n(\tilde\omega)Z(\tilde\omega)=\Bigl( H^u_{T^n\tilde\omega\wedge \omega', T^n\tilde\omega }   H^s_{\omega', T^n\tilde\omega\wedge \omega'}\Bigr) Z(\omega'),
$$
which implies that
$$
Z(\tilde\omega)=[A_n(\tilde\omega)]^{-1}\Bigl( H^u_{T^n\tilde\omega\wedge \omega', T^n\tilde\omega }   H^s_{\omega', T^n\tilde\omega\wedge \omega'}\Bigr) Z(\omega').
$$
It remains to note that
$$
Z(\omega)=\Bigl( H^u_{\omega\wedge \tilde \omega, \omega }   H^s_{\tilde\omega, \omega\wedge \tilde\omega}\Bigr) Z(\tilde\omega).
$$

\section{End  of the proof of Theorem~\ref{thm2}}

 Let  $E\in {\frak U}(A,\mu)$. We must show  that   $E\in {\frak U}(\tilde A,\tilde \mu)$.

Assume first   that   $0<|\Delta_p(E)|<2$  for some $T$- periodic point  $p\in \Omega$.
By the symbol $n_p$,   we  denote  the period of $p$.  We  also set
$$
L(A,p)=\lim_{n\to\infty}\frac1n \ln\bigl(\|A_n(p)\|\bigr).
$$
It is easy to see  that $Z(p)$,  viewed as a  set  of complex  numbers,   is   not real.  In fact,  $Z(p)$ consists  of one point  $(a+i\sqrt{4-(\Delta_p(E))^2})/b$,
where $a$  and $b\neq 0$  are the two elements of  the first  row of the matrix $A_{n_p}(p)$.
Since,   for any periodic point $q\in \Omega$,  the set  $Z(q)$ is     the image of $Z(p)$   under an ${\rm SL}(2,{\Bbb R})$  transformation,   $Z(q)$ is not real and  consists  of one point.   Therefore,   for any periodic point $q\in \Omega$,   the matrix $A_{n_q}(q)$
has two  complex  eigenvalues   that belong to  the unit  circle.  
The latter observation leads  to the conclusion that $E\in \sigma(H_q)$ and
\begin{equation}\label{L=0}
L( A,q)=0\qquad \text{for all periodic points}\quad q\in \Omega.
\end{equation}
In particular, $L(A,q)=L(\tilde A,q)=0$  for all  periodic points  that belong to $\tilde \Omega$.

Now we  use  the following  result  proved in a much more general setting by Kalinin (see  Theorem 1.4  in \cite{K}).
\begin{proposition}\label{approxL} Let $A$  be locally constant  on $\tilde \Omega$.
 Then  for each $\delta>0$ there is a periodic point $q\in \tilde \Omega$  such that $|L(\tilde A,q)-L(\tilde A,\tilde \mu)|<\delta$.
\end{proposition}
Combining Proposition~\ref{approxL}  with the  equality \eqref{L=0}, we obtain that $$L(\tilde A,\tilde \mu)=0.$$
Thus, $E\in {\frak U}(\tilde A,\tilde \mu)$.

Now assume   that $|\Delta_p(E)|\in\{0,2\}$  for all  $p\in {\rm Per}(T)$.  Then $|\Delta_p(E)|\in\{0,2\}$  for all  $p\in {\rm Per}(\tilde T)$.
In particular, this implies that all eigenvalues of $A_{n_p}$  belong to the unit  circle  and,  hence,  $L(\tilde A,p)=0$  for any periodic $p\in \tilde \Omega$.  Thus,  we infer  from Proposition~\ref{approxL} that $L(\tilde A,\tilde\mu)=0.$

The proof is complete.  $\,\,\Box$

\begin{corollary}
A point $E\in {\frak L}(A,\mu)$  is unremovable  from ${\frak L}(A,\mu)$  if and only if    the point $E$ belongs to   the   spectrum of   $H_p$   for each $p\in {\rm Per}(T)$.
\end{corollary}

The zeros of the Lyapunov exponent  that  belong to the  set
\[
{\frak S}(T,\mu)= \bigcup_{p\in {\rm Per}(T)} \bigl\{E\in {\Bbb R}:\,\,\Delta_p(E)\in (-2,0)\cup(0,2)\bigr\}\]
 have  simple and interesting  properties   described  in  the following statement.

\begin{corollary}\label{thm1} Let $V$ be a  real-valued   locally  constant function on $\Omega$.   Let $T:\, \Omega\to\Omega$ be  a subshift of finite type.
Assume that $\mu$ is a $T$-ergodic  measure on $\Omega$ that has a local product  structure and the property ${\rm supp}(\mu)=\Omega$.
Then   for any   subshift $\tilde T:\, \tilde\Omega\to\tilde \Omega$   of $T$
  and  any 
$\tilde T$-ergodic measure $\tilde \mu$ on  $\tilde \Omega\subset \Omega$,
$$
{\frak S}(T,\mu)\cap  {\frak L}(A,\mu)\,\subseteq \,\,{\frak S}(T,\mu)\cap  {\frak L}(\tilde A,\tilde \mu),
$$
where $\tilde A$ is the restriction of $A$  to $\tilde \Omega$.
\end{corollary}

\section{Proof of Theorem~\ref{thm3}}

Observe that under  the assumptions of  Theorem~\ref{thm3},
$$
N(\tilde A,\tilde \mu)\geq N(A,\mu).
$$

First  consider   the case   where ${\frak U}(A,\mu)={\frak U}(\tilde A,\tilde \mu)$   while  ${\frak S}(\tilde T,\tilde \mu) \subset{\frak S}(T,\mu)$ is a proper  inclusion.  Then  the inequality
\begin{equation}\label{<E}
{\mathcal J}(\tilde A,\tilde \mu)\leq {\mathcal J}( A,\mu)
\end{equation}
may be established by the  means of  Calculus.   
Indeed,  since  $N(\tilde A,\tilde \mu)\geq N(A,\mu)$, we only   need to show  that  the derivative   of ${\mathcal J}( A,\mu)$  with respect to $x=|(E_{j_0},E_{j_0+1})\cap {\frak S}(T,\mu)|$ is  positive,  provided
$|(E_0,E_{l+1})\cap {\frak S}(T,\mu)|$  is   viewed as   the linear function  $\tau-x$,  where $\tau=E_{j_0+1}-E_{j_0}+\sum_{j\neq j_0}|(E_{j},E_{j+1})\cap {\frak S}(T,\mu)|$.  Put differently,
we  must show that the derivative of  
$$
\psi(x)=\frac{x}{\lambda}\ln\Bigl(\frac{x}{\lambda}\Bigr)+\frac{\tau-x}{\lambda}\ln\Bigl(\frac{\tau-x}{N(A,\mu)\cdot \lambda}\Bigr)
$$
is positive.  

The direct  computation shows  that
$$
\psi'(x)=\frac{1}{\lambda}\ln\Bigl(\frac{N(A,\mu)\cdot x}{\tau-x}\Bigr).
$$
Thus, we  infer  that  $\psi'(x)>0$  from the finequality $(N(A,\mu)+1)x>\tau$.

It remains to prove \eqref{<E}  in the case   where  ${\frak U}(A,\mu)\subset {\frak U}(\tilde A,\tilde \mu)$    is a proper  inclusion. 
In the case ${\frak U}(A,\mu)={\frak U}(\tilde A,\tilde \mu)$,   the  quantity ${\mathcal J}( A,\mu)$ was  an increasing function   of   ${\frak S}(T,\mu)$.    Therefore,  it  is enough   to  consider  the case   where
${\frak S}(T,\mu)={\frak S}(\tilde T,\tilde \mu)$.  

Suppose  ${\frak U}(\tilde A,\tilde \mu)$  consists  of the points $\tilde E_1<\tilde E_2<\dots <\tilde E_r$.  Define $\tilde E_0=E_0$  and $\tilde E_{r+1}=E_{l+1}$.
Then each interval $[E_j,E_{j+1})$  is the   union of a finite collection  of intervals  $[\tilde E_k , \tilde E_{k+1})$:
$$
[E_j,E_{j+1})=\bigcup_{k= k_0(j)}^{\tilde k(j)}
[\tilde E_k , \tilde E_{k+1}).
$$
Therefore,
\[\begin{split}
|[E_j,E_{j+1})\cap {\frak S}(T,\mu)|\ln\Bigl(|[E_j,E_{j+1})\cap {\frak S}(T,\mu)|\Bigr)\geq  \\
\sum_{k=k_0(j)}^{\tilde k(j)} |[E_k,E_{k+1})\cap {\frak S}(T,\mu)|\ln\Bigl(|[E_k,E_{k+1})\cap {\frak S}(T,\mu)|\Bigr),
\end{split}
\]
which implies \eqref{<E}.

\section{The monotonicity is not strict}

Here  we give an example of a subshift   for which  $ \Omega$ is a proper  subset of  ${\mathcal A}^{\Bbb Z}$,   and yet  ${\frak U}(A,\mu)\,=\emptyset$.

Assume that $V$ depends  only on   the zero-coordinate  $\omega_0$  of $\omega$.
Then   all  stable and unstable holonomies are    identity operators.  Consequently,  if $m_\omega$  is a continuous  disintegration of  an  su-state,  then
$$
m_{\omega}=m_{\omega'}\quad \text{whenever}\quad \omega_0=\omega'_0.
$$
But then  the equality  
$$
A(\omega)m_\omega=m_{T\omega}
$$  implies 
that $m_{T\omega'}=m_{T\omega}$  whenever $ \omega_0=\omega'_0$.    

Let us now  give   a  condition  that  makes  the latter equality  impossible.
For each $j_0\in{\mathcal A}$,   define   the set
\begin{equation}\label{Dj}
D_{j_0}=\{j\in{\mathcal A}: \,\, \exists \omega\in \Omega\, \text{such that}\,\, \omega_0=j_0,\, \omega_1=j\}.
\end{equation} 
Suppose that  for each pair  of  symbols $j$  and $j'$ in ${\mathcal A}$,  there is an ordered   collection of  letters $j_1, j_2,\dots,j_k$,
such that $j\in D_{j_1}$ and $j'\in D_{j_k}$, while $D_{j_n}\cap D_{ j_{n+1}}\neq \emptyset$  for all  $1\leq n\leq k-1$.  Then $m_\omega$ is constant on $\Omega$.
This would imply that $Z(\omega)$  is constant on $\Omega$, which   would turn  the relation
$
A^E(\omega)\cdot Z(\omega)=Z(T\omega)
$
into   the  equality 
$$
A^E(\omega)\cdot Z(\omega)=Z(\omega).
$$  This   cannot be true 
in the case  where $V(\omega)$  takes at least two  different  values.  The obtained  contradiction  shows  that  there is no su-state  for the cocycle $A^E$,
which  implies that  the  Lyapunov  exponent is positive  for every $E\in {\Bbb R}$.

\begin{theorem}
Assume that $V$ depends  only on   the zero-coordinate  $\omega_0$  of $\omega$ and takes  at least two different values on $\Omega$.
Let the sets $D_{j_0}$ be defined  by \eqref{Dj}.  Suppose that  for each pair  of  symbols $j$  and $j'$ in ${\mathcal A}$,  there is an ordered   collection of  letters $j_1, j_2,\dots,j_k$,
such that $j\in D_{j_1}$ and $j'\in D_{j_k}$, while $D_{j_n}\cap D_{ j_{n+1}}\neq \emptyset$  for all  $1\leq n\leq k-1$.  Finally,  assume that $\mu$ is a $T$-ergodic  measure on $\Omega$ that  has a local product  structure and   the property ${\rm supp}(\mu)=\Omega$.Then the Lyapunov exponent is positive for each  $E\in{\Bbb R}$:
$$
L(A^E,\mu)>0.
$$
\end{theorem}

\end{document}